\documentclass[a4paper,fleqn]{cas-dc}
\usepackage{soul}
\usepackage{amsmath}
\usepackage[version=4]{mhchem}
\usepackage[numbers,sort&compress]{natbib}
\usepackage{tabularx}
\usepackage[flushleft]{threeparttable}
\usepackage[switch]{lineno}
\def\tsc#1{\csdef{#1}{\textsc{\lowercase{#1}}\xspace}}
\tsc{WGM}
\tsc{QE}
\tsc{EP}
\tsc{PMS}
\tsc{BEC}
\tsc{DE}

\begin{document}
\begin{sloppypar}

\let\WriteBookmarks\relax
\def\floatpagepagefraction{1}
\def\textpagefraction{.001}
\shorttitle{Nanoextraction from a flow of a highly diluted solution for much-improved sensitivity in offline chemical detection and quantification}
\shortauthors{Hongyan Wu et~al.}

\title [mode = title]{Nanoextraction from a flow of a highly diluted solution for much-improved sensitivity in offline chemical detection and quantification}

\author[1]{Hongyan Wu}
\credit{Writing- Original Draft, Methodology, Investigation, Validation}
\address[1]{Department of Chemical and Materials Engineering, University of Alberta, Alberta T6G 1H9, Canada}

\author[1]{Quynh Nhu Le}
\credit{Methodology, Investigation}

\author[1]{Binglin Zeng}
\credit{Writing- Original Draft}

\author[1,2]{Xuehua Zhang}
\cormark[1]
\address[2]{Physics of Fluids Group, Max Planck Center Twente for Complex Fluid Dynamics, JM
Burgers Center for Fluid Dynamics, Mesa+, Department of Science and Technology,
University of Twente, Enschede 7522 NB, The Netherlands}
\credit{Conceptualization of this study, Supervision, Writing - Review \& Editing, Funding acquisition, Project administration}
\cortext[cor1]{Corresponding author: xuehua.zhang@ualberta.ca (X. Zhang)}


\begin{abstract}
Preconcentration of the target compound is a critical step that ensures the accuracy of the subsequent chemical analysis. In this work, we present a straightforward yet effective liquid-liquid extraction approach based on surface nanodroplets (i.e., nanoextraction) for offline analysis of highly diluted sample solutions. The extraction and sample collection were streamlined in a 3-m microcapillary tube. The concentration of the target analyte in surface nanodroplets was significantly increased compared to the concentration in the sample solution, reaching several orders of magnitudes. A limit of detection (LOD)  was decreased by a factor of $\sim 10^3$ for an organic model compound in Fourier-transform infrared spectroscopy (FTIR) measurements and $\sim 10^5$ for a model fluorescent dye in fluorescence detection. The quantitative analysis of the organic compound was also achieved in a wide concentration region from $10^{-3}$ M to $10^{-4}$ M. The total volume of surface nanodroplets can be manipulated to further enhance extraction efficiency, according to the principle that governs droplet formation by solvent exchange. Additionally, our method exhibited significantly improved sensitivity compared to traditional dispersive liquid-liquid microextraction (DLLME). The LOD of the fluorescent dye and the organic model compound obtained with DLLME was 3 orders of magnitude and 20 times higher than the LOD achieved through nanoextraction approach. The nanoextraction developed in this work can be applied to preconcentrate multi-compounds from river water samples, without clear interference from each other. This can further extend its applicability for the detection and quantification of target analytes in complex aqueous samples by common analytical instruments.  
\end{abstract}


\maketitle

\section{Introduction}
Dispersive liquid-liquid microextraction (DLLME) is widely applied in analytical chemistry and plays an important role in analyte preconcentration.\cite{vinas2014dispersive,zgola2011dispersive,leong2014beyond,santana2021deep,zgola2010application} In the DLLME process, extractant microdroplets are spontaneously formed via the "ouzo effect" and stably dispersed in a mixture of disperser solvent and aqueous sample solution.\cite{lohse2020physicochemical,qian2019surface,hosseini2020ultrasound,jain2023simple} The large surface-to-volume ratio of microdroplets greatly increases the mass transfer rate of the target analyte across the droplet surface and hence improves the extraction efficiency in a short time.\cite{theberge2010microdroplets,xia2019recent,thakur2021functional} The dispersed droplets with concentrated analytes can be collected for analysis after separation with the assistance of a centrifuge. There are still some limitations in DLLME, including the dilution effect from a large amount of disperser solvent and the need for an equipment-assistant separation procedure.\cite{ojeda2011separation,rezaee2006determination,rezaee2010evolution} Surface nanodroplets with femtoliter volume can serve as an extraction phase to overcome several drawbacks of DLLME. However, it remains unexplored how surface nanodroplets can be leveraged for ultrasensitive detection.

A simple solution based approach, namely solvent exchange,  has been recently developed for producing surface nanodroplets.  A large quantity of surface nanodroplets can be formed in a fluidic system. By a solvent exchange process, a one-phase ternary solution consisting of a good solvent, a poor solvent, and an oil (i.e. extractant) is displaced by the flow of the poor solvent.\cite{you2021surface,dyett2018growth} The oversaturation of oil created by the solubility gap between the good and poor solvent leads to nucleation and growth of oil nanodroplets on the hydrophobic substrate inside the flow chamber.\cite{zhang2015formation} The final droplet size can be simply manipulated by the flow conditions (i.e., flow rate and flow geometry) and solution properties (i.e., solution chemistry and concentration).\cite{zhang2015formation,lu2015solvent,lu2016influence} Moreover, multicomponent surface nanodroplets can be formed by using solutions containing these components. The composition of the droplets is tuned by the oversaturation level of each component in the good and poor solvent, as reflected in the three-phase diagram.\cite{li2018formation,you2021tuning}                                         

Surface nanodroplets of water-insoluble extractants can directly enrich target analytes from the continuous flow-in aqueous solution. This process is referred to nanoextraction,\cite{li2019functional} due to the small sizes of the surface nanodroplet.  The extract rate of surface nanodroplets is exceptionally high, due to their huge surface-to-volume ratio. The extraction rate of a droplet can be determined as\cite{wei2022interfacial,jeannot1996solvent}

\begin{equation}
\frac{dC_o}{dt}=\frac{A}{V_o}{\beta}({p}{C_{aq}}-C_o),
\label{e0}
\end{equation}
where $C_o$ is the concentration of target analyte in the oil droplet at time $t$, $C_{aq}$ is the analyte concentration in the aqueous sample at time $t$, $A$ is the interfacial area of droplet, $V_o$ is the droplet volume, $\beta$ is the mass transfer coefficient related to the droplet liquid (i.e. extractant), and $p$ is the partition coefficient at the equilibrium, which is defined by ${C_{o}}/{C_w}$. Although eq. 1 neglects the effect from the contact angle of surface nanodroplets, the relation indicates that for a certain extraction phase and $C_{aq}$, increasing the surface-to-volume ratio of droplets is an effective way to maximize the extraction rate.

Extraction from a large volume solution is applicable to enrichment techniques such as solid-phase extraction or filtration\cite{schulze2017assessment,enamorado2015levels}. This approach is particularly useful for extracting target analytes from highly diluted solutions because a substantial amount of the sample is necessary to ensure that a sufficient quantity of analytes is enriched. A prime example of this scenario is when analytes are extracted from wastewater or urine samples. However, the large volume extraction is not typically employed in techniques such as DLLME or microdroplets in a microfluidic system. On the other hand, nanoextraction presents a promising technique for enriching analytes from large volume sample solutions. This technique relies on the stability of surface nanodroplets in an external flow, which is due to the pinning effect from the substrate.

After the formation of surface nanodroplets by solvent exchange, in-situ analysis can be immediately performed by surface-sensitive analytical instruments (e.g., fluorescence microscope and confocal Raman spectroscopy) during the extraction process.\cite{li2019functional,yu2016large} Moreover, we recently demonstrated that the liquid-liquid nanoextraction inside a capillary tube enables continuous enrichment of target analytes, and subsequent droplet collection for ex-situ detection.\cite{li2022surface} The collection of surface nanodroplets after extraction can be simply achieved by injecting a stream of air into the capillary tube. However, it remains unclear how to maximize the extraction performance of surface nanodroplets to further improve the sensitivity in chemical analysis. Moreover, quantitative analysis of collected nanodroplets by other common analytical instruments (e.g., FTIR) has not been investigated.

In this work, we employed surface nanodroplets inside a capillary tube to extract and preconcentrate analytes from highly diluted aqueous solutions to reach high sensitivity in offline analysis. The extraction performance was demonstrated by analyzing the enriched model compounds in collected droplets via most commonly used analytic instruments, including FTIR, UV-vis spectroscopy, and fluorescence microscope. The total droplet volume was controlled to maximize extraction efficiency, based on the principle that governs droplet formation by solvent exchange. The developed nanoextraction approach can significantly improve the sensitivity of common analytical instruments, offering a versatile technique for offline chemical detection and quantification.

\section{Materials and methods}

\subsection{Chemicals and materials}
1-Octanol (99 \%, Fisher Scientific) was used as the droplet liquid. Ethanol (histological grade) was selected as the good solvent for the formation of surface nanodroplets. Triclosan (98 \%, TCI America), nonanoic acid (97\%, Fisher Scientific) and Nile red (99 \%, Acros Organics) were served as the analytes for nanoextraction. Water (18.2 M$\Omega$) was obtained from a Milli-Q purification unit (MilliporeSigma). River water was collected from the North Saskatchewan River in Edmonton. The river water was filtered in advance to remove particles and impurities and then stored at 4 $^{\circ}$C.

A 3 m Teflon capillary tube (MilliporeSigma) was used for the formation of surface nanodroplets and subsequent extraction. The inner diameter and outer diameter of the tube are 0.8 mm and 1.58 mm, respectively. The tube was thoroughly rinsed by injecting ethanol and water into it before use.

\begin{figure*}[htbp]
\centering
 \includegraphics[height=7cm]{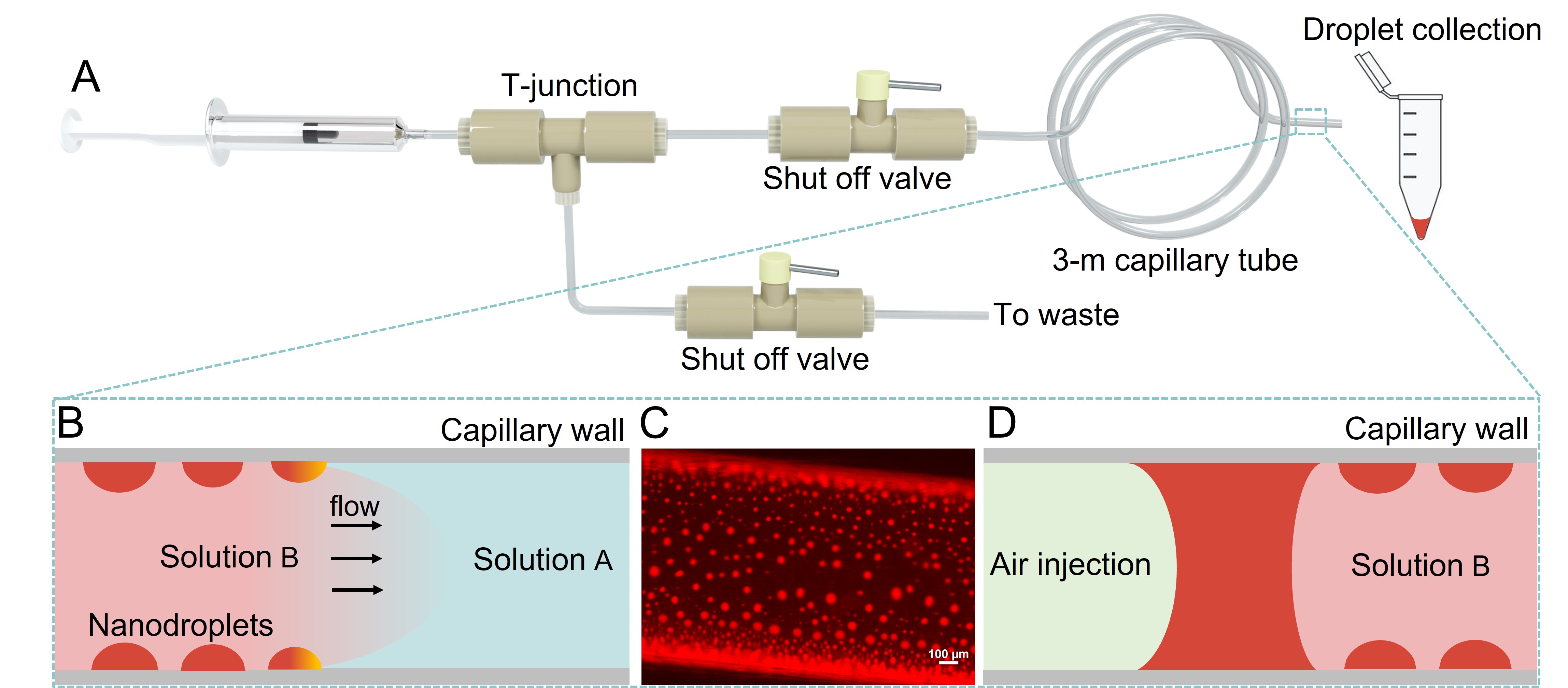}
  \caption{(A)Schematic showing the experimental setup for the formation, extraction, and collection of surface nanodroplets. (B) A sketch showing the droplet growth on the inner surface of the capillary tube by the solvent exchange, and subsequent extraction of analytes from solution B into surface droplets. (C) A fluorescent image showing the octanol surface droplets pinned on the inner wall of the tube. (D) A sketch showing the collection process of surface droplets by ﬂowing a stream of air into the capillary tube. }
\label{fgr:Setup}
\end{figure*}

\subsection{Surface nanodroplets formation and collection}

Surface nanodroplets were produced on the inner hydrophobic wall of the capillary tube by solvent exchange, as shown in Fig.\ref{fgr:Setup}A. A T-junction and a shut-off valve were used to connect the capillary tube and syringe. An additional shut-off valve was utilized to exclude trapped air in solutions and guarantee the mixing of two liquids. 

In the solvent exchange process, solution A and B were sequentially injected into the capillary tube with a controlled flow rate by a  syringe pump (New Era,
NE-1000). Solution A was prepared by adding 2 vol\% to 4.5 vol\% octanol into a mixture of water/ethanol (50 vol \%/50 vol \%). Solution B was an aqueous sample doped with target analytes and saturated octanol. The saturated octanol was used to prevent droplet dissolution in the flow of solution B. As sketched in Fig. \ref{fgr:Setup}B, surface nanodroplets of octanol formed at the mixing front of solution A and B. Immediately after the formation of octanol droplets on the inner surface of the capillary tube, the analyte dissolved in the aqueous sample was extracted and concentrated into surface nanodroplets. Fig. \ref{fgr:Setup}C shows a fluorescent image of surface nanodroplets with enriched Nile red on the inner surface of the tube (Nikon H600L).   
After the extraction was completed, a stream of air was injected into the capillary tube to collect surface nanodroplets. The air stream was controlled at a low flow rate of 3 mL/h, which pushed surface nanodroplets to accumulate at the air/water (solution B) interface and merge into a slug (Fig. \ref{fgr:Setup}D), until all droplets were expelled through the outlet of the tube. The final volume of collected octanol droplets was $\geq$ 1.5 $\mu$L, sufficient for the next step analysis.

\subsection{DLLME technique}

The sample solution was injected dropwise into a mixture of extraction solvent (octanol) and disperser solvent (ethanol) under constant magnetic stirring. The mixture was stirred for an additional 2 minutes to ensure complete extraction. Next, the cloudy emulsion was centrifuged for 10 min at 4000 rpm to separate the microdroplets. After this step, the floated octanol droplets with enriched analytes were collected using a microsyringe for analysis. Table 1 displays the six compositions of the ternary solution utilized for DLLME.   

\begin{table*}[htbp]
 \begin{threeparttable}
 \caption{Compositions of the ternary solution for DLLME.}
 \begin{tabular*}{1\textwidth}{@{\extracolsep{\fill}}cccccc}
\hline
Composition & Octanol ($\mu$L) & Ethanol (mL) & Water (mL) \\
\hline
A1          & 60           & 3            & 7          \\
B1          & 60           & 2            & 8          \\
C1          & 60           & 1            & 9          \\
A2          & 100          & 3            & 7          \\
B2          & 100          & 2            & 8          \\
C2          & 100          & 1            & 9 \\       
\hline
\end{tabular*}
\end{threeparttable}
\end{table*}

\subsection{Analytical methods and instruments}
A ﬂuorescent microscope (Nikon H600L) was used to confirm the enrichment of the fluorescent dye (Nile red) in collected octanol droplets. The collected octanol droplet was deposited on a glass slide and its fluorescence intensity was measured using green laser light to excite Nile red. The exposure time was kept constant at 1 s.

A Fourier-transform infrared spectroscopy (FTIR) spectrometer (Nicolet 6700, Thermo Fisher Scientific) coupled with a Smart Performer console was used to detect and quantify the nonanoic acid with a spectral resolution of 4 cm$^{-1}$. The number of scans was 64. The background spectrum of ambient air was subtracted to eliminate the signal contribution from instrument and environment. In addition, to quantitatively determine the analyte concentrations in the collected octanol droplets, we calibrate the absorbance from the analyte in standard octanol solution dissolved with analytes based on the Beer-Lambert law \cite{wagner2010use}.

An ultraviolet-visible (UV-vis) spectrometer (Nanodrop 2000c, Thermo Fisher) was employed to analyze the triclosan in collected octanol droplets. The UV-vis spectra of pure octanol was subtracted to obtain the absorbance peak of the extracted triclosan. 

\section{Results and discussion}
\subsection{Nanoextraction and IR detection}
Firstly, we demonstrate the capability of our approach to be integrated with infrared spectra measurement for the efficient detection and accurate quantification of concentrated target analytes in collected octanol droplets. To illustrate the principle, nonanoic acid was selected as the model compound due to its characteristic mid-IR spectral features, which can be readily identified and fingerprinted.\cite{sharma2021spectral, li2019hybrid} The sample volume and flow rate were controlled at 10 mL and 30 mL/h in the solvent exchange process. The initial octanol concentration in solution A remained at 2.5 vol\%.

\begin{figure*}[htbp]
\centering
 \includegraphics[height=5cm]{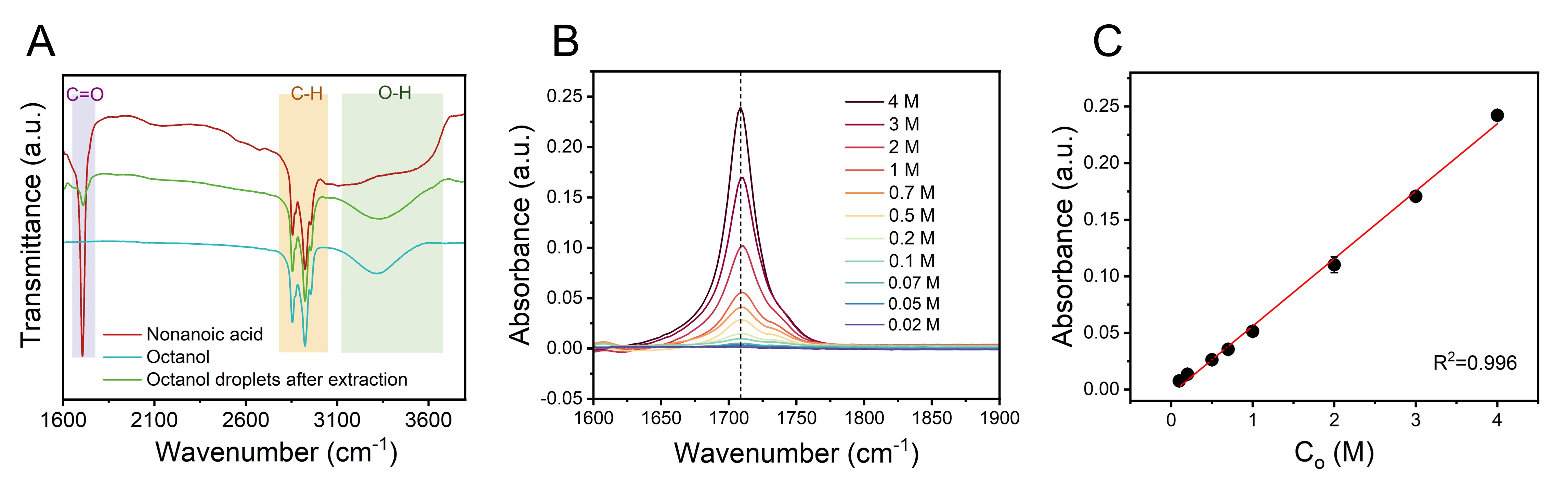}
  \caption{(A) FTIR spectra of pure octanol, pure nonanoic acid, and collected octanol droplets after extraction of $10^{-3}$ M nonanoic acid from aqueous solution. (B) FTIR absorbance spectrum of nonanoic acid in octanol standard solutions. The concentrations of nonanoic acid are 0.02 M-4 M.  (C) IR absorbance at 1709 cm$^{-1}$ as a function of nonanoic acid concentration in octanol standard solutions $C_o$. The error bars (standard deviation) are from three replicates. }
\label{fgr:Fig2}
\end{figure*}

Fig. \ref{fgr:Fig2}A shows the transmittance IR spectra of octanol, nonanoic acid, and collected octanol droplets after extraction from the aqueous sample containing $10^{-3}$ M nonanoic acid. The broad absorbance band at 3100-3600 cm$^{-1}$ is attributed to the O-H bond stretching.\cite{liu2021organic} The peaks at 2800-3100 cm$^{-1}$ are assigned to C-H bond stretching.\cite{liu2021organic} Those two bonds present in both nonanoic acid and octanol. The peak at 1709 cm$^{-1}$ reflects C=O bond stretching, which only appears in nonanoic acid.\cite{qian2020one} Therefore, this peak can be utilized as the characteristic peak of nonanoic acid, allowing for its distinguishment from the octanol background and enabling the subsequent quantification of its concentration.

The IR absorbance spectra at 1709 cm$^{-1}$ of nonanoic acid in octanol standard solution from 0.02 M to 4 M are shown in Fig. \ref{fgr:Fig2}B. The LOD of nonanoic acid without extraction was found to be 0.05 M, as no peak was observed at a lower concentration of 0.02 M. The IR signal exhibits a linear relationship at the concentration range 0.1 M-4 M with the variation coefficient (R$^2$) of 0.996 as shown in Fig. \ref{fgr:Fig2}C. According to the Beer-Lambert law,\cite{swinehart1962beer,schuttlefield2008atr} this linear correlation between IR absorbance and analyte concentrations in standard solution ($C_o$) can be used to quantify the concentrations of the extracted analyte in collected octanol droplets.      

\begin{figure*}[htbp]
\centering
 \includegraphics[height=14cm]{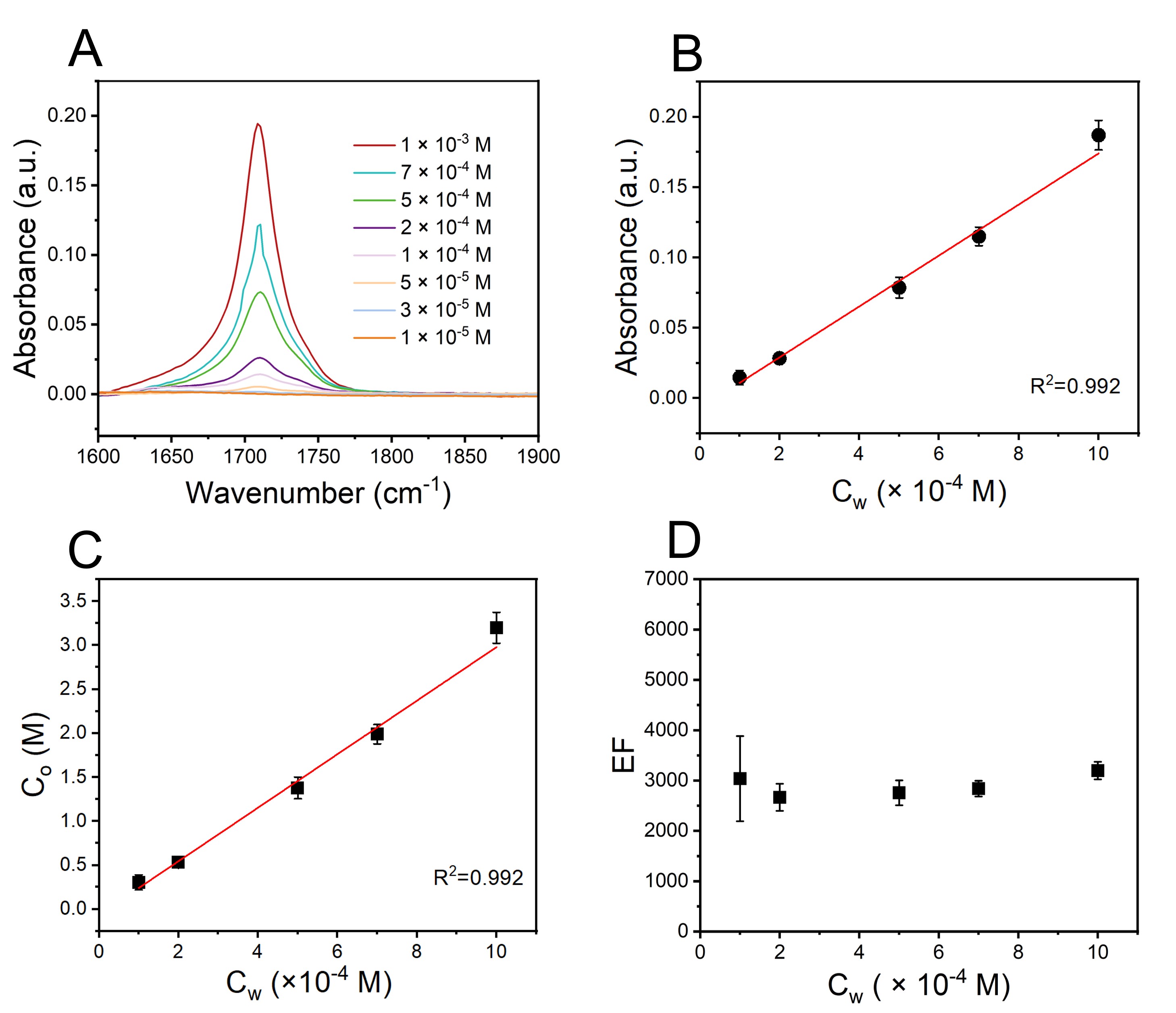}
  \caption{ (A) FTIR absorbance spectrum of collected octanol droplets. The droplets were collected after extraction from samples with nonanoic acid concentrations $C_w$ at $10^{-3}$-$10^{-5}$ M. (B) IR absorbance at 1709 cm$^{-1}$ as a function of $C_w$. The error bars (standard deviation) are from three replicates. (C) $C_o$ as a function of $C_w$. (D) Enrichment factor (EF) as a function of $C_w$. }
\label{fgr:Fig3}
\end{figure*}

Fig. \ref{fgr:Fig3}A shows the IR absorbance spectra of collected octanol droplets containing enriched nonanoic acid. The absorption peak at 1709 cm$^{-1}$ gradually decreases with the decrease in initial concentrations of nonanoic acid in aqueous sample solutions ($C_w$). With our nanoextraction method, a lower detection limit of $3\times10^{-5}$ was obtained. 
Compared to the LOD of 0.02 M without nanoextraction, the sensitivity is improved by a factor of $\sim 10^3$. Such a high enrichment efficiency can be attributed to two main processes. First, the huge surface-to-volume ratio and excellent stability of surface nanodroplets enable fast and continuous extraction of nonanoic acid across the droplet interface. Additionally, the enrichment was directly driven by the partition difference of the hydrophobic nonanoic acid between octanol and water ($K_{ow}$=3.52).\cite{medina2015discovery} According to eq.1, a higher $K_{ow}$ leads to a higher mass transfer rate. 

A linear relationship was found between the absorbance at 1709 cm$^{-1}$ and $C_w$ with a $R^2$ of 0.992 as shown in Fig. \ref{fgr:Fig3}B. By correlating the absorbance intensity in Fig. \ref{fgr:Fig2}C and Fig. \ref{fgr:Fig3}B, the relationship between $C_o$ and $C_w$ can be established (Fig. \ref{fgr:Fig3}C). This correlation allows for the quantification of the actual concentration of nonanoic acid in octanol droplets. The extraction efficiency of surface nanodroplets was further evaluated by calculating the enrichment factor (EF=$C_o$/$C_w$) from Fig. \ref{fgr:Fig3}C,\cite{mogaddam2021organic} where $C_o$ and $C_w$ represent the concentration of extracted analyte in collected octanol droplets and the initial analyte concentration in aqueous sample solutions, respectively.
 Fig. \ref{fgr:Fig3}D plots the EF as a function of $C_w$.
 We found that the EF kept around 3000 when $C_w$ ranged from $10^{-4}$ M to $10^{-3}$ M, indicating that the concentrations of nonanoic acid in octanol nanodroplets were increased by $\sim 10^3$ times with our nanoextration approach. 

\begin{figure*}[htbp]
\centering
 \includegraphics[height=5.5cm]{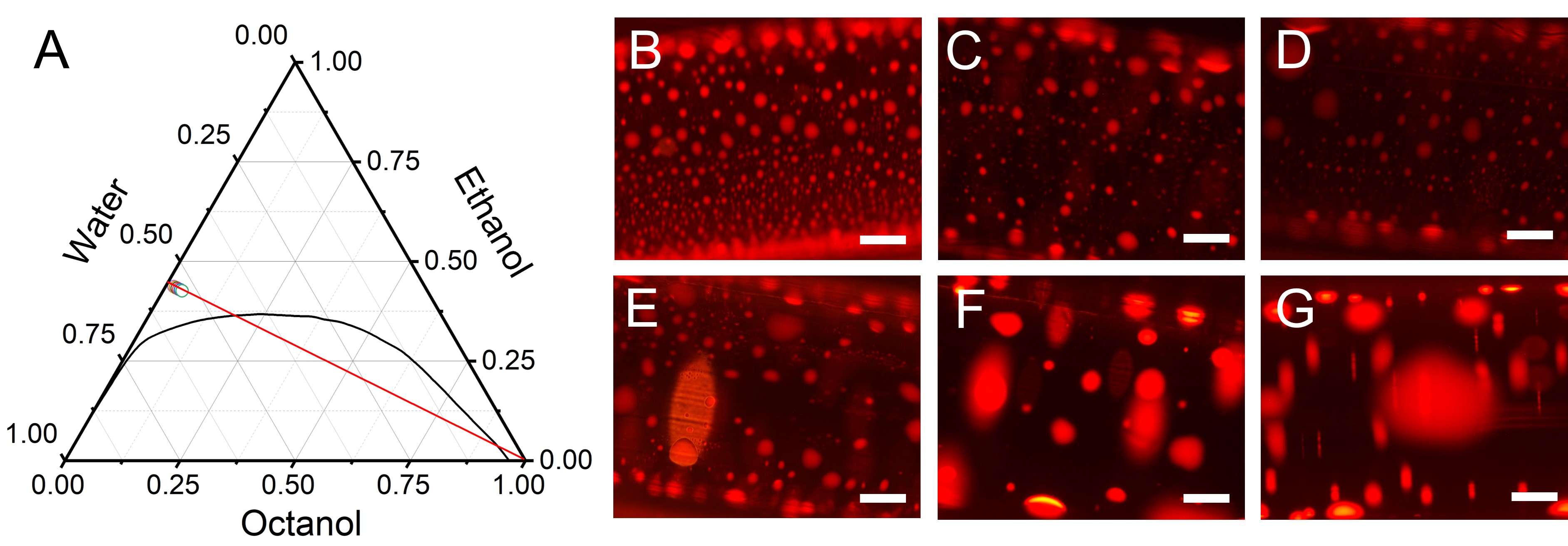}
  \caption{(A) Ternary phase diagram of octanol, ethanol, and water in wt\%.\cite{lopian2016morphologies} Six different initial conditions of solution A for the formation of octanol surface nanodroplets (colored circles).  Corresponding optical images of octanol surface droplets at initial octanol concentrations of (B) 2 vol\%, (C)2.5 vol\%, (D) 3 vol\%, (E) 3.5 vol\%, (F)4 vol\%, and (G) 4.5 vol\% in solution A. Scale bar: 200 $\mu$m}
\label{Fig4}
\end{figure*}

\subsection{Effects from total droplet volume on the extraction}
In solvent exchange process, the final volume of surface nanodroplets is formulated by the initial concentration of droplet liquid in solution A.\cite{bao2015highly} In order to assess the impact of droplet volume on extraction efficiency, the octanol content in solution A was varied from 2 vol\% to 4.5 vol\% while the ratio of ethanol to water remained constant, as shown in the ternary phase diagram (Fig. \ref{Fig4}A). The $C_w$ of nonanoic acid was remained at $10^{-3}$ M, and the sample volume and flow rate were controlled at 7 mL and 30 mL/h.

Figure. \ref{Fig4}B-G show the fluorescent images of octanol surface droplets at initial octanol concentrations from 2 vol\% to 4.5 vol\%. The size of octanol surface nanodroplets increases monotonically with an increase in initial octanol concentrations. Similar observations were reported in our previous work.\cite{lu2016influence} Lu et al. experimentally and theoretically revealed that the radius (R) of surface nanodroplets is proportional to the square root of the oversaturation area (A), namely 
\begin{equation}
R \sim A^{1/2}
\label{e7}
\end{equation}
where A is defined as the integrated region between the dilution path and the phase boundary, reflecting the oversaturation level of droplet liquid in the solvent exchange process.\cite {lu2016influence,li2019controlled}
Here we assume that the contact angle of surface droplets is constant, thus we can use the $R$ to characterize the droplet volume ($Vol$) and obtain $Vol \sim R^3$. Therefore, we can build the correlation between total droplet volume and oversaturation area, as demonstrated in eq. 3. 

\begin{equation}
Vol\sim R^3 \sim A^{3/2}
\label{7}
\end{equation}


\begin{figure*}[htbp]
\centering
 \includegraphics[height=14cm]{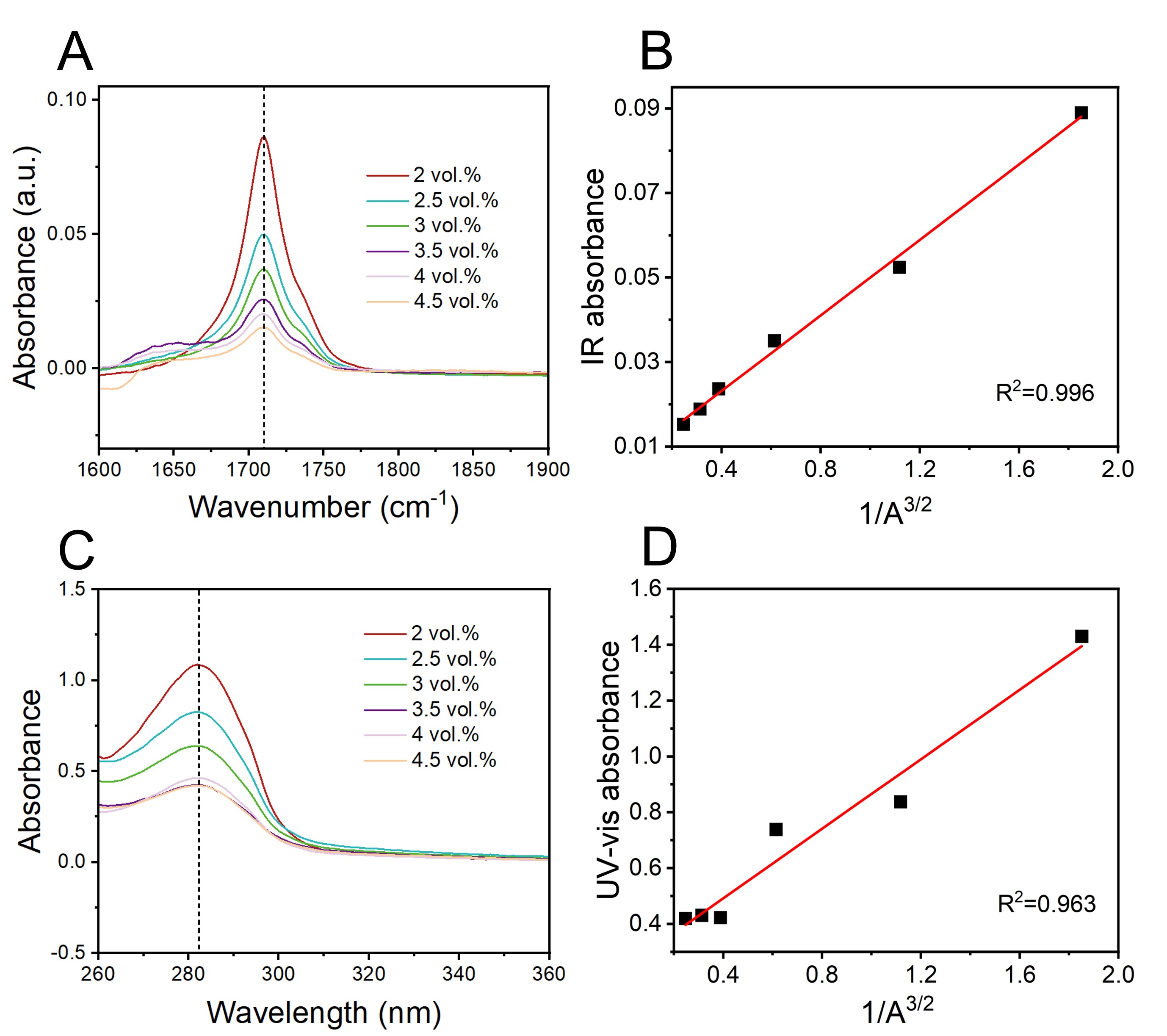}
  \caption{ (A) FTIR absorbance spectra of nonanoic acid in collected octanol droplets with the initial concentration of octanol in solution A varying from 2 vol\%-4.5vol\%. (B) IR absorbance at 1709 cm$^{-1}$ as the function of $A^{-3/2}$ and the fitting trendline. (C) UV-vis absorbance spectra of triclosan in collected octanol droplets with the initial concentration of octanol in solution A varying from 2 vol\%-4.5vol\%. (D) UV-vis absorbance at 283 nm as the function of $A^{-3/2}$ and the fitting trendline. }
\label{Fig5}
\end{figure*}

Fig. \ref{Fig5}A exhibits the IR absorbance of collected octanol droplets containing extracted nonanoic acid as a function of initial octanol contents in solution A. It was found that the absorbance monotonically increased with the decrease in octanol concentrations. The oversaturation area of different compositions of solution A was calculated based on the ternary phase diagram (Fig. S1). Then we build the correlation between absorbance (Abs) and oversaturation area ($A$), as shown in Fig. \ref{Fig5}B. It was found that the absorbance scaled as $Abs\propto A^{-3/2}$ with the oversaturation area. Based on the Beer-Lambert law, the absorbance at 1709 cm$^{-1}$ linearly correlated with the concentration of enriched nonanoic acid in octanol droplets ($Abs \propto C_o$), namely we can obtain $C_o\propto A^{-3/2}$. 

As a result, we can conclude that reducing the initial concentration of octanol in solution A results in a lower oversaturation level. This leads to a decrease in octanol droplet volume, which significantly imporves the preconcentration performance. 
While increasing the sample volume can also enhance the extraction efficiency, it requires more time to allow a larger volume of sample solution to flow into the capillary tube.\cite{li2022surface} 
Thus, adjusting the oil (extraction phase) concentration in solution A is a more effective way to achieve a high enrichment factor.
It is worth noting that the concentration of octanol in solution A cannot be further reduced below 2 vol\%, as the final volume of collected octanol droplets would then be too less (less than 1.5 $\mu$L) for detection. 




The dependence of extraction performance on the octanol concentration in Solution A was further verified by measuring the UV-vis absorbance of extracted triclosan, as shown in Fig. \ref{Fig5}C and D. 
Triclosan is a commonly used antibacterial drug and displays a strong absorbance peak at 283 nm in the UV-vis spectrum.\cite{yang2021solid} 
The absorbance at 283 nm was found to be monotonically increased with the decreasing of octanol concentration in solution A (Fig. \ref{Fig5}C). 
The scaling law $Abs\propto A^{-3/2}$ still holds true with the detection by a UV-vis spectrometer (Fig. \ref{Fig5}D). This result suggests that reducing the initial oil ratio in solution A is a general strategy to enhance the extraction efficiency of surface nanodroplets. 

\begin{figure*}[htbp]
\centering
 \includegraphics[height=7cm]{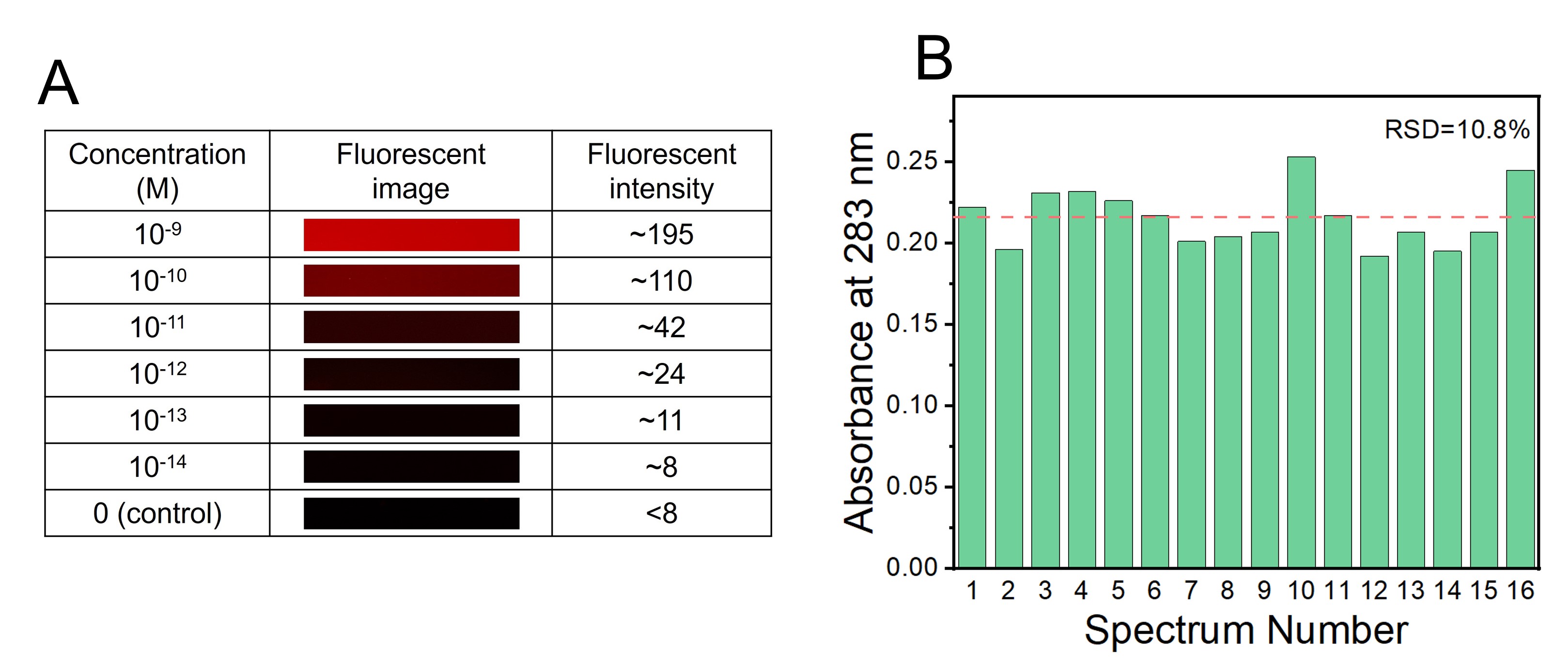}
  \caption{(A) Fluorescence images and intensities of collected octanol droplets. The droplets were collected after nanoextraction from samples with Nile red concentration at 10$^{-7}$ M-10$^{-14}$ M. (B) UV-vis absorbance at 283 nm versus spectrum number. }
\label{Fig6}
\end{figure*}

\subsection{Nanoextraction of multiple analytes}

The capability of surface nanodroplets to extract multiple compounds from the real sample solutions was also studied. Nile red ($K_{ow}$=5) and triclosan ($K_{ow}$=4.76) were spiked into river water to serve as solution B.\cite{bach1993biotechnology,kaur2019biomimetic} The concentrations of Nile red varied from 10$^{-7}$ M to 10$^{-14}$ M while the concentration of triclosan was fixed at 10$^{-7}$ M. 
The extraction performance of Nile red was examined by a fluorescent microscope, while the extraction performance of triclosan was analyzed by a UV-vis spectrometer. As shown in Fig. \ref{Fig6}A and B, both Nile red and triclosan were successfully enriched into surface nanodroplets, demonstrating that our nanoextraction technique has the capability to concentrate multiple compounds from the sample solution simultaneously. The LOD of Nile red reaches 10$^{-12}$ M, which is 10$^{5}$ times lower than the LOD of 10$^{-7}$ M without extraction (Fig. S2), demonstrating outstanding sensitivity by our approach. The small standard deviation of triclosan (10.8 \%) illustrates that the preconcentration effect is highly reproducible.

\subsection{Comparison of nanoextraction with classical DLLME}

\begin{figure*}[htbp]
\centering
 \includegraphics[height=11cm]{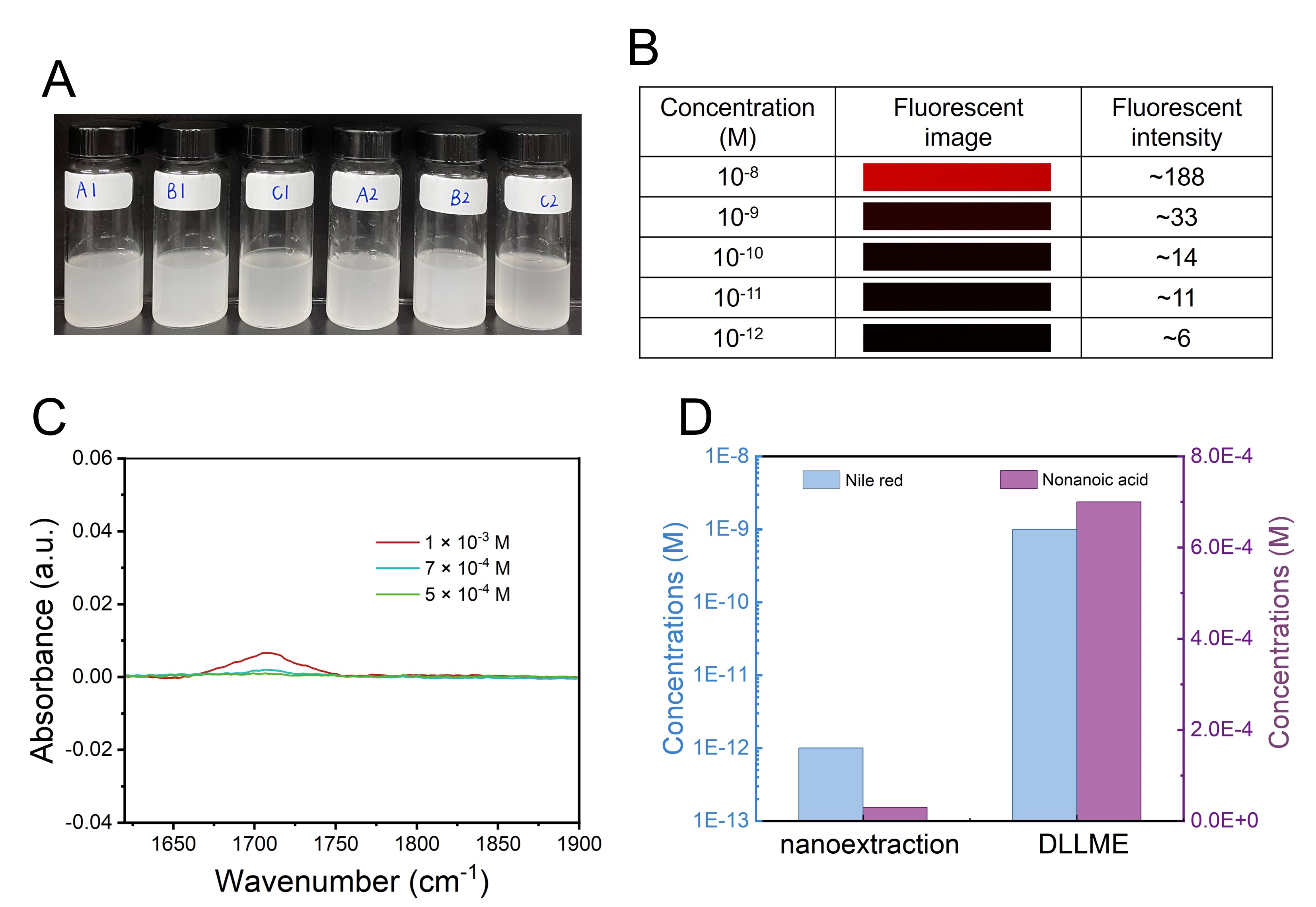}
  \caption{(A) A picture showing the cloudy emulsions of six different compositions of the ternary solutions for DLLME. The compositions are indicated in Table 1.
  (B) Fluorescence images and intensities of collected octanol droplets. The droplets were collected after DLLME from samples with Nile red concentration at 10$^{-8}$ M-10$^{-12}$ M. (C) FTIR absorbance spectrum of collected octanol droplets with enriched nonanoic acid after DLLME process. (D) Comparison of the LOD of Nile red and nonanoic acid obtained from DLLME and nanoextraction approaches.}
\label{DLLME LOD}
\end{figure*}

To better demonstrate the enhanced extraction efficiency achieved through our proposed method, we compared the sensitivity of this nanoextraction technique with classical DLLME approach. Based on six different compositions of the ternary solution (Table 1 and Fig. \ref{DLLME LOD}A), the final composition was optimized as A1, consisting of 60 $\mu$L of extraction solvent (octanol), 3 mL of disperser solvent (ethanol), and 7 mL of sample solution (Fig. S3). As depicted in Fig. \ref{DLLME LOD}B and C, the LOD of Nile red and nonanoic acid obtained with DLLME were found to be 10$^{-9}$ M and 7$\times10^{-4}$ M, respectively. Those values were 3 order of magnitude and 20 times higher than the LOD achieved through our nanoextraction approach, as shown in Fig. \ref{DLLME LOD}D.

The higher sensitivity of our nanoextraction technique can be mainly attributed to two main factors. Firstly, the extraction by surface nanodroplets takes places directly in the flow of the analyte solution through the capillary tube, hence eliminating the negative impact of a disperser solvent required in DLLME on the partition coefficient. Secondly, the smaller surface nanodroplets in nanoextraction provide a larger active surface area, leading to enhanced mass transfer rates compared to the microdroplets in DLLME. An additional advantage of the nanoextraction technique is the absence of separation equipment such as a centrifuge for droplet collection.    

\section{Conclusions}
In summary, we developed the nanoextraction approach based on surface nanodroplets inside a capillary tube for preconcentrating analytes from the continuous flow-in sample solutions. Surface nanodroplets with enriched analytes were accumulated into a liquid bridge by gently blowing air into the tube and collected for ex-situ analysis by common analytical tools including FTIR, UV-vis spectroscopy, and fluorescence microscope. With FTIR, the concentrations of the model compound nonanoic acid in octanol nanodroplets were found to be increased by a factor of $\sim 10^3$ compared to the initial concentrations in the aqueous sample. For fluorescence detection, the sensitivity was improved by a factor of $\sim 10^5$. The total volume of octanol surface nanodroplets can be reduced to greatly enhance the extraction efficiency. The capability of surface nanodroplets to synchronously
extract multi-compound samples containing Nile red and triclosan was also demonstrated. This approach is most advantageous compared to the classical DLLME for eliminating the adverse effect of the disperser solvent on the partition coefficient, improving enrichment factors, and hence sensitivity of the chemical analysis from highly diluted sample solutions.  The developed nanoextraction technique can be potentially applied for analyte preconcentration and detection in the areas of food control, environmental monitoring, or biomedical diagnosis.

\printcredits

\section*{Declaration of Competing Interest}

The authors declare that they have no known competing financial interests or personal relationships that could have appeared to influence the work reported in this paper.

\section*{Acknowledgments}
The authors acknowledge funding support from the Natural Sciences and Engineering Research Council of Canada (NSERC)\textendash Discovery project, Alliance-Alberta Innovates Advanced program, and the Canada Research Chairs program. H.Y.W acknowledges support from China Scholarship Council (No. 202106450020).

\appendix
\section*{Appendix A. Supplementary material}

Supplementary data associated with this article can be found in a separate file. 

\bibliographystyle{elsarticle-num}



\end{sloppypar}

\end{document}